\input harvmac


\def\np#1#2#3{Nucl. Phys. {\bf B#1} (#2) #3}
\def\pl#1#2#3{Phys. Lett. {\bf #1B} (#2) #3}
\def\prl#1#2#3{Phys. Rev. Lett. {\bf #1} (#2) #3}
\def\prd#1#2#3{Phys. Rev. {\bf D#1} (#2) #3}

\def\jhep#1#2#3{JHEP {\bf#1}(#2) #3}
\def\jmp#1#2#3{J. Math Phys. {\bf #1} (#2) #3}

\def\ijmp#1#2#3{Int.~J.~Mod.~Phys. {\bf #1} (#2) #3}
\def\atmp#1#2#3{Adv.~Theor.~Math.~Phys.{\bf #1} (#2) #3}

\def\IB{\relax\hbox{$\inbar\kern-.3em{\rm B}$}}
\def\IC{\relax\hbox{$\inbar\kern-.3em{\rm C}$}}
\def\ID{\relax\hbox{$\inbar\kern-.3em{\rm D}$}}
\def\IE{\relax\hbox{$\inbar\kern-.3em{\rm E}$}}
\def\IF{\relax\hbox{$\inbar\kern-.3em{\rm F}$}}
\def\IG{\relax\hbox{$\inbar\kern-.3em{\rm G}$}}
\def\IGa{\relax\hbox{${\rm I}\kern-.18em\Gamma$}}
\def\IH{\relax{\rm I\kern-.18em H}}
\def\IK{\relax{\rm I\kern-.18em K}}
\def\IL{\relax{\rm I\kern-.18em L}}
\def\IP{\relax{\rm I\kern-.18em P}}
\def\IR{\relax{\rm I\kern-.18em R}}
\def\IZ{\relax\ifmmode\mathchoice{
\hbox{\cmss Z\kern-.4em Z}}
{\hbox{\cmss Z\kern-.4em Z}}
{\lower.9pt\hbox{\cmsss Z\kern-.4em Z}}
{\lower1.2pt\hbox{\cmsss Z\kern-.4em Z}}
\else{\cmss Z\kern-.4em Z}\fi}
\def\II{\relax{\rm I\kern-.18em I}}

\def\ndt{{\noindent}}
\def\ads#1{{\bf AdS}_{#1}}


\def\CL{{\cal L}}

\def\CN{{\cal N}}

\def\CR{{\cal R}}
\def\CS{{\cal S}}

\def\p{\partial}
\def\pb{\bar{\partial}}

\def\p{\partial}
\def\pb{\bar{\partial}}

\def\Tr{{\rm Tr}}

\def\vol{{\rm vol}}

\def\Det{{\rm Det}}


\def\inbar{\,\vrule height1.5ex width.4pt depth0pt}

\font\cmss=cmss10 \font\cmsss=cmss10 at 7pt


\def\b{{\beta}}

\def\g{{\gamma}}

\def\vf{{\varphi}}
\def\m{{\mu}}

\def\t{{\theta}}
\def\o{{\omega}}

\def\N{{\bf N}}


\lref\tasi{J.~Polchinski, ``TASI Lectures on D-Branes'',
  {\tt hep-th/9611050}}

\lref\dgmorb{M.R. Douglas, B.R. Greene and D.R. Morrison,
``Orbifold Resolution by D-branes'', \np{505}{1997}{84},
  {\tt hep-th/9704151}}

\lref\dougegs{M.R. Douglas, ``Enhanced Gauge Symmetry
in M(atrix) theory'', \jhep{007}{1997}{004},
  {\tt hep-th/9612126}}

\lref\jmorb{C.V. Johnson and R.C. Myers, ``Aspects of
Type $\II$B Theory on ALE Spaces'', \prd{55}{1997}{6382},
  {\tt hep-th/9610140}}

\lref\gipol{E.G.~Gimon and J.~Polchinski, ``Consistency Conditions for
Orientifolds and D Manifolds'', \prd{54}{1996}{1667},
  {\tt hep-th/9601038}}

\lref\polten{J.~Polchinski, ``Tensors from K3 Orientifolds'',
\prd{55}{1997}{6423}, {\tt hep-th/9606165}}

\lref\kmvgeom{S.~Katz, P.~Mayr and C.~Vafa,
  ``Mirror Symmetry and
  Exact Solution of 4D $\CN=2$ Gauge Theories -- I'',
  {\tt hep-th/9706110}}

\lref\branegeom{H.~Ooguri and C.~Vafa, ``Geometry of $\CN=1$ dualities
in four dimensions'', \np{500}{1997}{62}; {\tt hep-th/9702180}}

\lref\absorb{S. Gubser and I. Klebanov,``Absorption by Branes
and Schwinger Terms in the World Volume Theory,''
\pl{413}{1997}{41}}

\lref\KW{I.R. Klebanov and E. Witten, ``Superconformal Field Theory on
Threebranes at a Calabi-Yau Singularity,'' 
{\it Nucl. Phys.} {\bf B536} (1998) 199, {\tt hep-th/9807080}.}

\lref\GK{
S.S. Gubser and I.R. Klebanov, ``Baryons and Domain Walls in an
{\cal N}=1 Superconformal Gauge Theory,'' {\it Phys. Rev.} {\bf D58} 
(1998) 125025, hep-th/9808075.}

\lref\MP{D.R. Morrison and M.R. Plesser,
``Non-Spherical Horizons, I,'' hep-th/9810201.}

\lref\SG{S.S. Gubser, ``Non-conformal examples of AdS/CFT,''
hep-th/9910117.}

\lref\Gir{L. Girardello, M. Petrini, M. Porrati, A. Zaffaroni,
``Novel Local CFT and Exact Results on Perturbations of N=4 
Super Yang Mills from AdS Dynamics,''
hep-th/9810126.}

\lref\DZ{J. Distler and F. Zamora,
``Nonsupersymmetric conformal field theories from
stable anti-de Sitter spaces,'' hep-th/9810206;
``Chiral Symmetry Breaking in the AdS/CFT Correspondence,''
hep-th/9911040.}

\lref\Freed{D.Z. Freedman, S.S. Gubser, K. Pilch, N.P. Warner,
``Renormalization Group Flows from Holography--Supersymmetry and a c-Theorem,''
hep-th/9904017.}

\lref\mfour{E.~Witten, ``Solutions of Four-dimensional
  Field Theories via M-theory'', \np{500}{1997}{3}; {\tt hep-th/9703166}}

\lref\aspegs{P.S. Aspinwall, ``Enhanced Gauge Symmetries and K3
Surfaces'', \pl{357}{1995}{329}, {\tt hep-th/9507012}}

\lref\bsvegs{M. Bershadsky, V. Sadov and C. Vafa, ``D-strings on
D-manifolds'',
\np{463}{1996}{398}}

\lref\dixon{L. Dixon, D. Friedan, E. Martinec and S. Shenker,
``The Conformal Field Theory of Orbifolds,''
\np{282}{1987}{13}}

\lref\br{M. Bershadsky and A. Radul,
``Conformal Field Theories with Additional $\IZ_N$ Symmetry,''
\ijmp{A2}{1987}{165}}

\lref\kwdiscrete{L.M. Krauss and F. Wilczek,
  ``Discrete Gauge Symmetries
in Continuum Theories'', \prl{62}{1989}{1221}}

\lref\reidrev{M. Reid, ``McKay correspondence'',
  {\tt alg-geom/9702016}}

\lref\suthree{W.M.~Fairbanks, T.~Fulton and W.H.~Klink, ``Finite and
Disconnected Subgroups of $SU(3)$ and their Application to the
Elementary Particle Spectrum'', \jmp{5}{1964}{1038}}
\lref\sosix{W.~Plesken and M.~Pohst, Math. Comp. {\bf 31} (1977) 552}
\lref\bkv{M.~Bershadsky, Z.~Kakushadze and  C.~Vafa, 
``String expansion as large N expansion of gauge theories",
\np {523} {1998} {59}, {\tt hep-th/9803076}}
\lref\bj{M.~Bershadsky and A.~Johansen, {\tt hep-th/9803249}}

\lref\KT{I.~R.~Klebanov and A.~A.~Tseytlin, ``D-Branes and
Dual Gauge Theories in Type 0 Strings,''
\np{546}{1999}{155}, {\tt hep-th/9811035}}

\lref\KTc{I.~R.~Klebanov and A.~A.~Tseytlin,
  ``Non-supersymmetric CFT from Type 0 String Theory,''
\jhep{9903}{1999}{015}, {\tt hep-th/9901101}
}

\lref\DM{M.~Douglas and G.~Moore,
``D-branes, quivers, and ALE instantons,'' {\tt hep-th/9603167}}

\lref\DH {L.~Dixon and J.~Harvey,
``String theories in ten dimensions without
space-time supersymmetry", \np{274}{1986}{93} \semi
N.~Seiberg and E.~Witten,
``Spin structures in string theory", \np{276}{1986}{272}\semi
C.~Thorn, remarks at the workshop
``Superstring Theories and
 the Mathematical Structure of Infinite-Dimensional Lie Algebras'',
 Santa Fe Institute, November 1985}

\lref\sw{N.~Seiberg and E.~Witten,
``Monopole Condensation and Confinement 
In N=2 Supersymmetric Yang-Mills Theory,''
\np{426}{1994}{19}}

\lref\vafa{
S. Katz, P. Mayr, C. Vafa, ``Mirror symmetry and 
Exact Solution of 4D ${\CN}=2$ gauge theories I'',
hep-th/9706110, 
 {\it Adv.Theor.Math.Phys. 1} (1998) 53-114}

\lref\berg{O.~Bergman and M.~Gaberdiel, ``A Non-supersymmetric Open
String Theory and S-Duality,'' \np{499}{1997}{183},
{\tt hep-th/9701137}}

\lref\jthroat{J.~Maldacena,
``The Large N limit of superconformal field theories and
  supergravity,'' \atmp{2}{1998}{231},
{\tt  hep-th/9711200} }

\lref\gkp{S.S.~Gubser, I.R. Klebanov, and A.M. Polyakov,
  ``Gauge theory correlators from noncritical string theory,''
  \pl{428}{1998}{105}, {\tt hep-th/9802109}}

\lref\EW{E.~Witten, ``Anti-de Sitter space and holography,''
\atmp{2}{1998}{253},
{\tt hep-th/9802150}}

\lref\AP{A.M.~Polyakov, ``The Wall of the Cave,''
{\tt hep-th/9809057}}

\lref\KS{S.~Kachru and E.~Silverstein, ``4d conformal field theories
and strings on orbifolds,''  \prl{80}{1998}{4855},
{{\tt hep-th/9802183}}.}
\lref\LNV{A.~Lawrence, N.~Nekrasov and C.~Vafa, ``On conformal field
theories in four dimensions,'' \np{533}{1998}{199},
{{\tt hep-th/9803015}}}

\lref\GNS{S.~Gubser, N.~Nekrasov, S.~Shatashvili,
``Generalized Conifolds and 4d N=1 SCFT,'' hep-th/9811230.}

\lref\NS{N. Nekrasov and S. Shatashvili,
``On non-supersymmetric CFT in four dimensions,''
{\tt hep-th/9902110.}, L.~Okun Festschrift, North-Holland,
in press}

\lref\KNS{I.R. Klebanov, N. Nekrasov and S. Shatashvili,
``An Orbifold of Type 0B Strings and Non-supersymmetric Gauge Theories,''
hep-th/9909109}

\lref\JM{J. Minahan, ``Glueball Mass Spectra and Other Issues for
Supergravity Duals of QCD Models,'' {\tt hep-th/9811156}}

\lref\JMnew {J. Minahan, ``Asymptotic Freedom and Confinement from Type 0
String Theory,'' {\tt hep-th/9902074}}

\lref\KTnew{I.R. Klebanov and A.A. Tseytlin, ``Asymptotic Freedom and
Infrared Behavior in the Type 0 String Approach to Gauge Theory,''
\np{547}{1999}{143}, {\tt hep-th/9812089} }

\lref\Blum{R. Blumenhagen, A. Font and D. Lust,
``Non-Supersymmetric Gauge Theories from
D-Branes in Type 0 String Theory,''
{\tt hep-th/9906101.}}

\lref\BCR{M. Bill\' o, B. Craps and  F. Roose, ``On D-branes in
Type 0 String Theory,'' {\tt hep-th/9902196.}
}

\lref\Zar{K. Zarembo, ``Coleman-Weinberg Mechanism and Interaction of
D3-branes in Type 0 String Theory, {\tt hep-th/9901106}}

\lref\ceresole{A. Ceresole, G. Dall'Agata, R. D'Auria, and S. Ferrara,
``Spectrum of Type IIB Supergravity on $\ads{5}\times {\bf T}^{1,1}$: 
Predictions
On ${\CN}=1$ SCFT's,'' hep-th/9905226.}

\lref\diac{D.-E. Diaconescu, M. Douglas and J. Gomis, ``Fractional Branes
and Wrapped Branes,'' {\it JHEP} {\bf 02} (1998) 013.}

\lref\Kehag{
A. Kehagias, ``New Type IIB Vacua and Their F-Theory Interpretation,''
{{\tt hep-th/9805131}}.
}

\lref\SGub{S.S. Gubser, ``Einstein Manifolds and Conformal Field Theories,''
{\it Phys. Rev.} {\bf D59} (1999) 025006, hep-th/9807164. 
}

\lref\cd{
P.~Candelas and X.~de la Ossa, ``Comments on Conifolds,''
{\it Nucl. Phys.} {\bf B342} (1990) 246.}

\lref\Das{
K.~Dasgupta and S.~Mukhi, ``Brane Constructions, Fractional
Branes and Anti-de Sitter Domain Walls,'' hep-th/9904131.}

\lref\jpp{C.~Johnson, A.~Peet and J.~Polchinski, in preparation;
reported by C.~Johnson, talk at IAS.}

\Title{\vbox
{\baselineskip 10pt
\hbox{PUPT-1897}
\hbox{ITEP-TH-61/99}
\hbox{hep-th/9911096}
{\hbox{   }}}}
{\vbox{\vskip -30 true pt
\centerline {Gravity Duals of Fractional Branes}
\medskip
\centerline {and Logarithmic RG Flow}
\medskip
\vskip4pt }}
\vskip -20 true pt
\centerline{ Igor R.~Klebanov$^{1}$ and Nikita A.~Nekrasov$^{1,2}$ 
}
\smallskip\smallskip
\centerline{$^{1}$ \it Joseph Henry
Laboratories, Princeton University, Princeton, New Jersey 08544}
\centerline{$^{2}$ \it Institute for Theoretical and Experimental
Physics, 117259 Moscow, Russia}

\bigskip\bigskip
\centerline {\bf Abstract}
\baselineskip12pt
\noindent
\medskip
We study fractional branes in
${\CN}=2$ orbifold and ${\CN}=1$ conifold theories.
Placing a large number $N$ of regular D3-branes at the singularity
produces the dual ${\bf AdS}_5\times X^5$ geometry, and we describe
the fractional branes as small perturbations to this background.
For the orbifolds, $X^5={\bf S}^5/\Gamma$ and fractional D3-branes excite
complex scalars from the twisted sector which are localized on
the fixed circle of $X^5$. The resulting solutions are given by holomorphic
functions and the field-theoretic beta-function is simply reproduced.
For $N$ regular and $M$ fractional D3-branes at the conifold singularity
we find a non-conformal ${\cal N}=1$ supersymmetric $SU(N+M)\times SU(N)$
gauge theory. The dual Type $\II$B background is 
${\bf AdS}_5\times {\bf T}^{1,1}$
with NS-NS and R-R 2-form fields turned on. This dual description
reproduces the logarithmic flow of couplings found in the field theory.
\bigskip

\Date{11/99}

\noblackbox \baselineskip 15pt plus 2pt minus 2pt

\newsec{Introduction}

By now there exists an impressive body of evidence that Type
$\II$B strings on $\ads{5}\times X^5$ are dual to large $N$
strongly coupled 4-d conformal gauge theories, in the sense proposed
in \refs{\jthroat,\gkp,\EW}. Here $X^5$ are positively curved
5-d Einstein spaces whose simplest example ${\bf S}^5$ corresponds to
the ${\cal N}=4$ supersymmetric $SU(N)$ gauge theory.
One may also quotient this duality by a discrete subgroup $\Gamma$
of the $SU(4)$ R-symmetry \refs{\KS,\LNV}. The resulting backgrounds
with $X^5 = {\bf S}^5/\Gamma$ are dual to ``quiver gauge theories'' 
with gauge group $S( U(N)^n )$ and bifundamental matter \DM,
which describe D3-branes near orbifold singularities. 
In such orbifold theories, in addition to regular D-branes which can reside
on or off the orbifold fixed plane there are also ``fractional'' D-branes
pinned to the fixed plane \refs{\gipol,\dougegs}. 
Our goal in this paper is to consider
the effect of such fractional branes on the dual supergravity
background. There is good
motivation for studying this problem because, as we discuss below,
introduction of fractional branes
breaks the conformal invariance and introduces RG flow.

It has also been possible to construct dual gauge theories for
$X^5$ which are not locally ${\bf S}^5$. The simplest example is
$X^5= {\bf T}^{1,1}= (SU(2)\times SU(2))/U(1)$ which turns out to be dual
to an ${\CN}=1$ superconformal $SU(N)\times SU(N)$ gauge
theory with a quartic superpotential for bifundamental fields 
\refs{\KW,\MP}.
In this theory, which arises on D3-branes at the conifold singularity,
it is also possible to introduce fractional D-branes 
\refs{\GK,\Das}, and we
study their effects in this paper.

Having constructed the gravity duals of fixed-point theories, the
next logical step is to study the dual picture of the RG flow.
A natural setup for this problem is provided by the
supersymmetric flows connecting orbifold theories and the
(generalized) conifold theories. It is clear, though, that in order
to have a consistent picture one cannot restrict oneself to 
the $\Gamma$ invariant supergravity fields only,
and needs to add the massless fields
coming from the twisted sectors. This program was initiated
in \refs{\KW,\GNS}, and
in this paper we focus on the role of the twisted sector fields
in creating RG flows.
Considerable progress on supersymmetric RG flows in other situations
has also been made recently 
\refs{\Freed}
(for a review see \SG). For important
work done in studies of non-supersymmetric 
RG flows see  \refs{\Gir,\DZ}.

The basic picture common to all RG problems
is that the radial coordinate
$r$ of $\ads{5}$ defines the RG scale of the field theory, hence the scale
dependence of couplings may be read off from the radial dependence
of corresponding supergravity fields.

Since the RG flows of couplings in physically relevant gauge 
theories are logarithmic, an important problem is to find gravity
duals of logarithmic flows. Attempts in this direction have been made
in the context of Type 0B string theory \refs{\KT,\JM,\KTnew}.
This is an NSR string with
the non-chiral
GSO projection $(-1)^{F+\tilde F}=1$
which breaks all spacetime supersymmetry \DH\ (it is also a 
$(-1)^{F_s}$ orbifold of the Type $\II$B theory). 
Type 0B theory has two basic types of D3-branes, electric and magnetic, and
we will see that it is appropriate to call them the fractional branes.
If equal numbers of the electric and magnetic branes are stacked
parallel to each other, then we find on their world volume a
$U(N)\times U(N)$ 
gauge theory coupled to six adjoint scalars of the first $U(N)$,
six adjoint scalars of the second $U(N)$, and Weyl fermions in
bifundamental representations.
This theory is a ``regular''
${\IZ}_2$ orbifold of the ${\cal N}=4$ $U(2N)$ gauge
theory and hence is conformal in the planar limit \refs{\KTc,\NS}.
\foot{The ${\IZ}_2$ is generated by
$(-1)^{F_s}$, where $F_s$ is the fermion number,
together with conjugation by
$\gamma= \pmatrix{ I&  0\cr  0 & -I\cr}$
where $I$ is the $N \times N$ identity matrix. This orbifold
is called regular because $\gamma$ is traceless.}

In general, in orbifold theories there are as many types
of fractional
branes as there are nodes of the quiver diagram (i.e. the number of
gauge groups in the product), or the number of the irreducible
representations of the orbifold group $\Gamma$.
If one takes a collection of the fractional
branes of each type and the branes corresponding to the 
irreducible representation ${\CR}_i$ are taken $n_i = 
{\rm dim} {\CR}_i$ times then one gets a single
regular D-brane which can depart to the bulk. More specifically,
the charge of the fractional brane of ${\CR}_i$ type is
\eqn\brch{{\rm q}_i =  {{n_i}\over{\vert \Gamma \vert}}}
where $\vert\Gamma\vert$ is the order of the orbifold group.
It is well known that $\sum_i n_i {\rm q}_i = 1$.

The fractional branes act as sources for the twisted closed string
states of the orbifold theory.
In the Type 0B example discussed above the two $U(N)$ groups correspond
to the electric and the magnetic D3-branes, hence these are the
two types of fractional branes for this particular ${\IZ}_2$ orbifold.
Indeed, such branes have tadpoles for the twisted RR 4-form and for
the tachyon \refs{\KT,\berg}.
If we stack $N$ parallel electric branes only, then we find an
``irregular'' orbifold theory where the $\IZ_2$ action does not
act on the gauge indices. This gives $SU(N)$ gauge theory
coupled to six adjoint scalars, which is not a CFT \KT.

Equations satisfied by the gravity dual of this theory were derived
in \KT, and solved after some assumptions in \refs{\JM,\KTnew}.
The RG flow of the dilaton, which is related to the gauge coupling,
comes from the equation
\eqn\dflow{ \nabla^2 \phi = - {1\over 4\alpha'} T^2 e^{\phi/2}
\ .}
Since the tachyon field $T$ has a source $F_5^2$, it departs from zero
and causes the dilaton to depend on $r$. Assuming that $T$ approaches
a constant for large $r$ it was found that the RG flow is logarithmic in
the UV \refs{\JM,\KTnew}. While this scenario has a number of
uncertainties (due to the lack of detailed knowledge of the $T$-dependence
in the effective action) it suggests a mechanism for RG flow of couplings
in the dual gravity picture of fractional branes. In particular,
the presence of ``twisted'' fields sourced by the
fractional branes plays the crucial role.

The stack of electric D3-branes defines gauge theory coupled to six
scalars fields, but it is obviously of more interest to remove
the scalars and study the pure glue theory \AP.
One way of embedding it into string theory is to consider
a ${\IZ}_{2}$ orbifold of Type 0B by reflection of six coordinates
\KNS. The regular orbifold theory has
gauge group $U(N)^4$ coupled to a chiral field content:

\item{$\bullet$}
four quadruples
of bi-fundamental fermions
transforming in $({\bf N}_{\bf i}, {\overline {\bf N}}_{{\bf i}+1})$,
${\bf i} = 0,1,2,3\semi \quad 4 \equiv 0$,

\item{$\bullet$}
four sextets of bi-fundamental
scalars in
$({\bf N}_{\bf i}, {\overline {\bf N}}_{{\bf i}+2})$.

\ndt
Now there are 4 gauge groups in the product,
hence there should be 4 different types of fractional branes. If we stack
$N$ fractional D3-branes of the same type then we find pure glue
$U(N)$ gauge theory on their world volume \KNS.
This suggests that fractional branes may provide a link to string
duals of realistic gauge theories.\foot{It seems unclear, however,
whether the fractional D3-branes in the ${\IZ}_4$ theory
which are stuck to the fixed fourplane
exist, for their flux has nowhere to escape to.}

With this eventual goal in mind, in this paper we study gravity duals
of fractional branes in supersymmetric conifold and orbifold theories:
the SUSY removes some of the effective action uncertainties present
in the Type 0B case. To simplify matters further we consider theories
where $\beta$--functions for `t Hooft couplings 
$g_{\rm YM}^2 N$ are of order
$1/N$ rather than of order $1$. Such gauge theories occur on $M$
fractional D3-branes parallel to $N$ regular D3-branes, with $M$
held fixed in the large $N$ limit. In the simplest examples we
consider, the gauge group is then $SU(N+M)\times SU(N)$.

The large number $N$ of regular D3-branes produces the dual $\ads{5}\times
X^5$ background, and for our purposes we may ignore the back-reaction of
the $M$ fractional D3-branes on it. However, the fractional
branes act as sources for an extra set of fields, namely the
2-form potentials $B^{NSNS}$ and $B^{RR}$. The flux of these 2-forms
through a certain 2-cycle of $X^5$ (more precisely its
deviation from the value at the orbifold point)
defines the difference between $g_{\rm YM}^{-2}$
for the gauge groups factors. In the ${\CN}=2$ supersymmetric orbifold cases
the 2-cycles are collapsed, so that the twisted sector fields
corresponding to the 2-form fluxes are confined to 
$\ads{5}\times {\bf S}^1$
where ${\bf S}^1$ is the fixed circle of ${\bf S}^5/\Gamma$.
By studying the dependence of these twisted sector fields
on the $\ads{5}$ radial coordinate $r$
we find the logarithmic flow of the gauge couplings 
consistent with field theory expectations. In fact, the
twisted sector fields are given by holomorphic functions of
the complex variable $z= r e^{i\varphi}$ where $\varphi$ is the
${\bf S}^1$ coordinate (previously known solutions of 
${\cal N}=2$ gauge theories are also characterized by holomorphic
functions \refs{\sw,\mfour,\vafa}).\foot{
Supergravity duals of logarithmic RG flows in
${\cal N}=2$ gauge theories are being independently studied in
\jpp.}
Remarkably, in the limit we are studying all string-scale
corrections are small, so that the use of effective supergravity
equations is justified.

\newsec{ Fractional Branes on the Conifold.}

Let us start by reviewing what is known about regular D3-branes
at conical singularities. If a large number $N$ of D3-branes is placed
at the apex of a 6-d cone $Y^6$ with metric
\eqn\cone{ds_{\rm cone}^2 = d r^2 + r^2 ds_5^2\ , 
}
then the near-horizon region of the resulting 10-d geometry has
the metric 
\eqn\genmet{ ds^2 = R^2 \left [ 
{r^2\over R^4} (-dt^2 + dx_1^2 + dx_2^2 + dx_3^2) + {dr^2\over r^2}+
ds_5^2 \right ]\ ,\qquad\qquad R^4\sim g_s N (\alpha')^2\ .
}
This geometry is $\ads{5}\times X^5$ where 
$X^5$ is the base of the cone 
(if $Y_6$ is Ricci-flat then $X^5$ is a positively
curved Einstein space \refs{\Kehag,\KW}.
Type $\II$B theory on this background is conjectured to be dual to
the conformal limit of the gauge theory on $N$ D3-branes placed
at the apex \refs{\KW,\MP}. 

An example where such a duality
has been tested extensively is when $Y^6$ is the conifold, which is
a singular Calabi-Yau manifold
described in terms of complex variables
$w_1,\dots, w_4$ by the equation \cd
$$ \sum_{a=1}^4 w_a^2 = 0
\ .
$$
The base of this cone is 
${\bf T}^{1,1}= (SU(2)\times SU(2))/U(1)$
whose Einstein metric may be written down explicitly \cd,
\eqn\co{ d s_{X_5}^2=
{1\over 9} \left(d\psi + 
\cos \theta_1 d\phi_1+ \cos \theta_2 d\phi_2\right)^2+
{1\over 6} \sum_{a=1}^2 \left(
d\theta_a^2 + \sin^2\theta_a d\phi_a^2 \right)
\ .}
The ${\CN}=1$
superconformal field theory on $N$ regular D3-branes placed at the singularity
of the conifold has gauge group $SU(N)\times SU(N)$ and global
symmetry $SU(2)\times SU(2)\times U(1)$ \KW. The chiral superfields
$A_1$, $A_2$ transform as $(\N,\overline \N)$ and are a doublet of the first
$SU(2)$;
the chiral superfields
$B_1$, $B_2$ transform as $(\overline \N, \N)$ and are a doublet of the second
$SU(2)$. The R-charge of all four chiral superfields is $1/2$ and the
theory has an exactly marginal superpotential
$W=\epsilon^{ij} \epsilon^{kl}\Tr A_iB_kA_jB_l$.

$\II$B supergravity modes on $\ads{5}\times{\bf T}^{1,1}$ have been matched in
some detail with operators in this gauge theory whose dimensions
are of order $1$ in the large $N$ limit \refs{\SGub,\ceresole}. 
In addition,
string theory has heavy supersymmetric states obtained by wrapping
D3-branes over 3-cycles of ${\bf T}^{1,1}$. They have been shown \GK\ to
correspond to ``dibaryon'' operators whose dimensions grow as $3N/4$ 
(schematically, these operators have the form ${\Det} A$ or ${\Det} B$).
Further, one may consider a domain wall in $\ads{5}$ obtained by
wrapping a D5-brane over the 2-cycle of ${\bf T}^{1,1}$ (topologically,
${\bf T}^{1,1}$ is ${\bf S}^2\times {\bf S}^3$). 
If this domain wall is located
at $r=r_*$ then, by studying the behavior of wrapped D3-branes upon
crossing it, it was shown in \GK\ that for $r> r_*$ the gauge group
changes to $SU(N+1)\times SU(N)$.
Note that this is precisely the gauge theory expected on $N$
regular and one fractional D3-branes! Thus, a D5-brane wrapped over the
2-cycle is nothing but a fractional D3-brane placed at a definite
$r$.\foot{Similarly, a regular D3-brane serves as a domain wall between
$SU(N)\times SU(N)$ and $SU(N+1)\times SU(N+1)$ gauge theory. This
has a simple interpretation in terms of Higgsing the theory.}
The identification of a fractional D3-brane with a wrapped D5-brane
is consistent with the results of \refs{\jmorb,\dougegs,\diac,\Das}.

As shown in \GK, this suggests a construction of the Type $\II$B
dual for the ${\CN}=1$ $SU(N+M)\times SU(N)$ gauge theory.
In particular, the background has to contain $M$ units of R-R
3-form flux through the 3-cycle of ${\bf T}^{1,1}$:
\eqn\RRflux{\int_{C^3} H^{RR} = M\ .
}
If $M$ is fixed as $N\rightarrow \infty$ then the
back-reaction of $H^{RR}$ on the metric and the $F_5$ background
may be ignored to leading order in $N$. However, as we will show,
the background must also include the NS-NS 2-form potential
\eqn\NSback{ B^{NSNS} = e^{\phi} f(r) \, {\o}_2 \ ,
}
where ${\o}_2$ is the closed 2-form corresponding to the 2-cycle
which is dual to $C^3$,
\eqn\NSflux{\int_{C^2} {\o}_2 = 1\ .
}
The desired connection with the RG flow is due to the fact
that \refs{\LNV, \KW,\GK,\MP}
\eqn\NSf{{1\over g_1^2} - {1\over g_2^2} \sim 
e^{-\phi} \left (\int_{C^2} B^{NSNS} - {1\over 2} \right )}
where $g_1$ and $g_2$ are the gauge couplings for $SU(N+M)$ and
$SU(N)$ respectively. Therefore, the $f(r)$ in \NSback\
gives the dual supergravity definition of the scale dependence
of $ {1\over g_1^2} - {1\over g_2^2}$.

Before solving for $f(r)$ let us recall the $\beta$-function
calculation in field theory. There we have
\eqn\betafun{ \eqalign{ 
{d\over d {\rm log} (\Lambda/\mu)}
{1\over g_1^2} &\sim 3(N +M) - 2N (1- \gamma_A - \gamma_B)\ ,\cr
{d\over d {\rm log} (\Lambda/\mu)}
{1\over g_2^2} &\sim 3N - 2(N+M) (1-  \gamma_A - \gamma_B)\ ,}
}
where $\gamma$ are the anomalous dimensions of the fields $A_i$ and $B_j$.
For $M=0$ we find a fixed point with $\gamma_A=\gamma_B =-1/4$ 
which corresponds
to R-charge $1/2$. This is the superconformal
gauge theory dual to $\ads{5}\times {\bf T}^{1,1}$
with vanishing 2-form potentials. For $M\neq 0$ it is impossible
to make both beta functions vanish (even if
we allow the anomalous dimensions of $A$ and $B$ to be different)
and the theory undergoes logarithmic RG
flow:
\eqn\RGsolve{
{1\over g_1^2} - {1\over g_2^2} \sim M {\rm log} (\Lambda/\mu) [ 3 + 2 
(1-\gamma_A - \gamma_B)]
\ .
}
Near the fixed point we expect 
$\gamma_A+\gamma_B=-1/2$ plus small corrections,
hence the RHS gives $\sim M {\rm log} (\Lambda/\mu)$. 

Let us reproduce
this result in supergravity. 
We need the Type $\II$B SUGRA equations of motion involving
the 2-form gauge potentials. We will write these equations
in the $\ads{5}\times {\bf T}^{1,1}$ background with 
constant $\tau = C_0 + i e^{-\phi}$
(this is the ${\rm SL}_{2}({\IZ})$
covariant combination of the dilaton and the R-R scalar of the
Type $\II$B theory):\foot{
We will check later that this background has no corrections of order
$M/N$ which is the order at which we are working.}
\eqn\fieldeqs{ d ^\star G = i F_5 \wedge G \ .
}
$G$ is the complex $3$-form field strength,
\eqn\threef{ G = H^{RR} + {\tau} H^{NSNS}\ , }
which satisfies the Bianchi identity $d G=0$.
Note that the RHS of \fieldeqs\ originates from the Chern-Simons
term
\eqn\CS{ \int C_4 \wedge H^{RR} \wedge H^{NSNS} \ .
}

Since the fractional D3-brane (the wrapped D5-brane) creates
R-R 3-form flux through ${\bf T}^{1,1}$, $H^{RR}$ should be proportional
to the closed 3-form which was constructed in \GK,
\eqn\harmonic{
H^{RR} \sim M e^\psi \wedge
( e^{\theta_1} \wedge e^{\phi_1} - 
e^{\theta_2} \wedge e^{\phi_2} )
\ .}
Here we are using the basis 1-forms
\eqn\basis{ e^\psi =
{1\over 3} (d\psi + \sum_{i=1}^2 \cos \theta_i d\phi_i )\ ,
\qquad e^{\theta_i} = {1\over \sqrt 6} d\theta_i\ ,
\qquad e^{\phi_i} = {1\over \sqrt 6} \sin \theta_i d\phi_i\ .
}
In these coordinates, the closed form ${\o}_2$ which enters 
\NSback\ is given by
$$ {\o}_2 \sim
e^{\theta_1} \wedge e^{\phi_1} -
e^{\theta_2} \wedge e^{\phi_2}
\ ,
$$
so that
$$ e^{-\phi} H^{NSNS}\sim d f(r)  \wedge
( e^{\theta_1} \wedge e^{\phi_1} -
e^{\theta_2} \wedge e^{\phi_2} )\ .
$$ 
Since $F_5= {\vol} (\ads{5} ) + {\vol} ({\bf T}^{1,1})$, 
$F_5\wedge H^{NSNS}=0$. Let us set the R-R scalar
$C_0=0$. Then
we see that the real part of \fieldeqs\ is satisfied for all $f(r)$.
{}From the imaginary part we have
$$ {1\over r^3} {d\over dr} \left(r^5 {d\over dr} f\left(r\right)\right) 
\sim M \ ,
$$
which implies 
\eqn\result{ f(r) \sim M {\rm log} \ r \ .}

Quite remarkably, our solution of Type $\II$B SUGRA
equations has reproduced the field theoretic beta function
for ${1\over g_1^2 N} - {1\over g_2^2 N}$ to order $M/N$.
This establishes the gravity dual of the
logarithmic RG flow in the ${\CN}=1$ supersymmetric $SU(N+M)\times
SU(N)$ gauge theory on $N$ regular and $M$ fractional D3-branes placed
at the conifold singularity. 

It is not hard to see that there are
no other effects of order $M/N$: the back-reaction of the 3-form
field strengths on the metric, dilaton and the $F_5$ comes in at order
$(M/N)^2$. Consider, for instance, the dilaton
equation of motion:
$$ \nabla^2 \phi ={1\over 12} (e^\phi H_{RR}^2 - e^{-\phi} H_{NSNS}^2)
\ .
$$
Even without checking the relative normalizations of $H_{RR}^2$ and
$H_{NSNS}^2$, we can immediately see that the variation of 
$ e^{-\phi}/N$ is at most of order $(M/N)^2$.
In the field theory this quantity translates into
${1\over g_1^2 N} + {1\over g_2^2 N}$. The fact that there is no
$\beta$-function of order $M/N$ for this quantity 
agrees with the field theory RG equations provided that the 
sum of the anomalous
dimensions, $\gamma_A + \gamma_B$, has no corrections of order $M/N$. 
This is
a simple gravitational prediction about the gauge theory.
It is of further interest to study the order $(M/N)^2$ effects
on the background and compare them with field theory, but we postpone
these calculations for future work.

\newsec{ Fractional Branes in ${\CN}=2$ Orbifold Theories.}

In this section we will be concerned with
the ${\CN}=2$ supersymmetric orbifolds of Type $\II$B strings.
Before introducing D-branes these are backgrounds of the form
$\IR^{5,1}\times \IR^4/\Gamma$. 
To write down gravity duals of fractional
branes we follow the strategy used in the last section:
we first stack a large number $N$ of regular D3-branes creating the dual
$\ads{5}\times {\bf S}^5/\Gamma$ background and then introduce the fractional
branes as small perturbations on this background.
We find an important difference from the conifold case, however,
because ${\bf S}^5/\Gamma$ does not have any finite volume 2-cycles.
A blowup of this space produces 2-cycles but breaks the ${\cal N}=2$
supersymmetry. 
For this reason we work with the singular space
where the 2-cycles are collapsed (we will see that this turns
out to be simpler than the non-singular case discussed in the previous
section).

Thus we consider the Type $\II$B superstring
in the background ${\bf AdS}_5 \times{\bf S}^5/{\Gamma}$ where
$\Gamma$ is one of the A,D,E subgroups of $SU(2)$
whose action on ${\bf S}^5$ is induced from the standard action on
${\bf R}^4 \approx {\IC}^2$ times the trivial action on ${\IR}^2$.
Therefore,
we get the action of $\Gamma$ on ${\IR}^6$ having a fixed two-plane
$0 \times {\IR}^2$. This action descends to ${\bf S}^5$ and the fixed
plane becomes a fixed circle ${\bf S}^1$.

The metric on ${\bf S}^5/{\Gamma}$ reads:
\eqn\intm{ds_5^2 = d{\t}^2 + {\cos}^2{\t} d{\vf}^2
+ {\sin}^2{\t} ds_{{\bf S}^3/{\Gamma}}^2
\ ,}
and the fixed circle is at $\theta=0$.

String theory in this background has extra massless fields compared to 
the $\Gamma$-invariant fields of the ten dimensional $\II$B supergravity.
These fields are localized at the fixed 
 circle ${\bf S}^1$. In the paper \GNS\ the multiplets
of the five dimensional gauge supergravity these fields fall in were 
identified.

The simple meaning of these fields is the following: if one were
to blow up the fixed circle to obtain a smooth Einstein metric
then the topology
of the resulting fivefold is such that the non-contractable
two-spheres $C_i$ are supported. These spheres intersect each other according
to the Dynkin diagram of the corresponding A,D,E Lie group:
$$
C_i \cap C_j = a_{ij}
$$

Now, reducing the ten dimensional supergravity fields along these
cycles leads to new fields in six dimensions spanned by
${\bf S}^1 \times {\bf AdS}_5$. Of particular importance for
us are the NSNS and RR two-forms $B_{NSNS}$ and $B_{RR}$. 
They give rise to the scalars:
\eqn\sclrs{{\b}_{i}^{NSNS} = \int_{C_{i}} B_{NSNS}, \quad
{\b}_{i}^{RR} = \int_{C_{i}} B_{RR}}
These fields are present even if one did not perform the blowup:
they come from the twisted sector of the string theory.

We now proceed with writing the effective action for these
fields (here we have set the R-R scalar to zero):
\eqn\lgrng{\int_{{\bf S}^1 \times {\bf AdS}_5} \sqrt{g} g^{mn} a^{ij}
\left( e^{-\phi} {\p}_m {\b}_i^{NSNS} {\p}_n {\b}_{j}^{NSNS} +
e^{\phi}
{\p}_m {\b}_i^{RR} {\p}_n 
{\b}_{j}^{RR}\right ) + a^{ij}
F_{5} \wedge {\b}_i^{NSNS} \wedge d {\b}_j^{RR} 
}
The last term comes from the 10-d Chern-Simons coupling of the
form
$$
\int F_{5} \wedge B_{NSNS}
 \wedge d B_{RR}
\ .$$

Introduce the complex fields:
$$
{\g}_i = {\tau} {\b}_{i}^{NSNS} + {\b}_i^{RR}
\ .$$
These fields transform nicely under 
the ${\rm SL}_{2}({\IZ})$ group:
$$
\pmatrix{ a & b \cr c & d} 
\cdot 
\pmatrix{
{\g}_i \cr
\tau}
 = {1\over{c\tau +d}} 
\pmatrix{
{\g}_i \cr a\tau + b}
$$
The metric on the ${\bf S}^1 \times 
{\bf AdS}_5$ space is:
$$
ds^2 = R^2\left(  {{dr^2}\over{r^2}} + d{\vf}^2 \right) +
{{r^2}\over{R^2}} (-dt^2 + dx_1^2 + dx_2^2+ dx_3^2)\ ,
\qquad R^4 = 4\pi g_{s} N {\ell}_{s}^4\ ,
$$
where ${\vf}$ is the coordinate on the circle ${\bf S}^1$. 

Let us introduce the complex coordinate on 
the 
space transverse
to ${\IR}^{1,3}$: $z = r e^{i\vf}$. 
We are interested in fields which have no ${\IR}^{1,3}$
dependence. 
For these fields
the action \lgrng\ takes on 
the following 
form (assuming that $\tau$ is constant, otherwise
we get covariant derivatives instead of ordinary ones):
\eqn\lgrngi{S = \int r^5 a^{ij}
 {1\over{\tau_2}}
{ {\p} {\g}_i\over \partial \bar z}
{ {\p} {\bar\g}_{j} \over \partial z}
d r d {\vf} }
where $\tau_2 = e^{-\phi} = {\rm Im}{\tau}$.
This action is ${\rm SL}_2({\IZ})$ invariant.

{\sl Gauge couplings.} The theory on the boundary of the
${\bf AdS}_5$ space is the superconformal quiver gauge theory with 
the gauge group $SU(Nn_0) \times SU(Nn_1) \times
\ldots \times SU(Nn_r)$ where $n_i$ are the Dynkin indices
-- the dimensions of the irreps of $\Gamma$. $n_0 = 1$ is the
dimension of the trivial representation ${\CR}_0$.

The relation between the boundary values of ${\g}_i$ and 
the couplings of these gauge factors was shown in \LNV\ to be:
\eqn\cplngs{\eqalign{
& 
\tau_i = {\g}_i, \quad i = 1, \ldots, r \cr
& \tau_0 = \vert 
\Gamma \vert
 \tau
 - \sum_i n_i \tau_i \cr}}
where
$$
\tau_j = {{{\t}_j}\over{2\pi}} + {{4\pi i}\over{
g_j^2}}$$

{\sl Running of the dilaton.}
Now let us discuss the validity of our assumption that $\tau$
is constant.
The Lagrangian for the ${\vec x}$ independent
dilaton reads as follows:
\eqn\dltn{{\CL} = 
\int r^3 t^3 dr dt d{\vf} {1\over{{\tau}_2}} 
\left( r^2 {\p}_r {\tau} {\p}_r {\bar \tau}
+ (1 - t^2) {\p}_t {\tau} {\p}_t {\bar\tau} + {1\over{1-t^2}}
{\p}_{\vf} {\tau} {\p}_{\vf} {\bar\tau}
\right) }
where we introduced the notation $t= {\sin}{\t}$.
The fixed circle is at $t=0$. The fields ${\g}$
acts as sources for the $\tau$ equations of motion. 
Irrespectively of the precise form of \dltn\ the source
term in the dilaton equation of motion is:
\eqn\src{\sim {\delta}(t)  r^5 e^{\phi}
a^{ij}
({\p}{\g}_i {\pb} {\g}_j + 
{\p}{\bar \g}_i {\pb} {\bar \g}_j )\ ,
}
(this expression is valid for $C_0=0$, $\phi = {\rm const}$).
Thus, the source term 
vanishes for holomorphic ${\g}$ and the dilaton is allowed to
remain constant. This argument may suffer
from some subtleties in case of badly singular ${\g}_i$.

{\sl Fractional branes.} In string theory $B_{NSNS}, B_{RR}$ do not
have to be globally well-defined two-forms - they behave like gauge fields.
The same applies to the scalars ${\b}_i$ obtained by the reduction
of the $B$-fields. In particular, if
we add the $k$-th 
fractional threebrane (D5-brane wrapped over $C_k$) at
some point $z_{*}$ then the scalars
 ${\b}_j^{RR}$ pick up a shift when circled around its location:
$$
{\b}_j^{RR} \to {\b}_j^{RR} + a_{kj} 
$$
The brane being BPS does not spoil the holomorphicity of
the ${\g}$'s, hence we conclude that the solution must have a logarithmic
monodromy:
\eqn\lgrthm{{\g}_{j} \sim {{a^{kj}}\over{2\pi i
}} {\rm log} ( z - z_{*} ) + \ldots }

One may be concerned about the appearance of the logarithm because  
$\gamma_j$ is defined  on a torus with modular parameter $\tau$.
Luckily, $\tau_2={1\over g_s}$ is of order $N$ in the `t Hooft limit
and the periodicity in this direction may be ignored for large $N$
(this is because the $\beta$-functions for $\gamma_j$ are of order 1
for a finite number of fractional branes, so that the evolution of the 
couplings is relatively slow).  In any event, the theory with a
fractional brane may be regulated far in the
UV by adding a fractional anti-brane at large $r$:
$${\g}_{j} \sim {{a^{kj}}\over{2\pi i
}} [{\rm log} ( z - z_{*} ) - 
{\rm log} ( z - z_{\rm reg} ) ] + \ldots
$$
This way the theory becomes conformal again for
$r \gg |z_{\rm reg}| $.
On the other hand, the singularities of $\gamma_i$ at
the locations of fractional branes are presumably removed
by the instanton corrections which will be discussed below.

{\sl Absence of the dilaton running from the field theory
expectations.}
The fractional brane of the ${\CR}_i$ type affects the gauge theory in a
simple way: by changing the $SU(N n_i)$ factor into
$SU(Nn_i + 1)$ and not changing the rest.

Clearly this induces a non-trivial beta functions for all couplings
$\tau_j$, such that $j = i$ or $a_{ij} \neq 0$. In fact:
$$
{{d}\over{d {\rm log}\left({\Lambda}/{\m}\right)}} {\tau}_j =  2N n_j
-
\sum_{k  \neq i}  a_{jk}
N n_k + (Nn_i + 1)  a_{ij} = - a_{ij} 
$$
$$
{d\over{d{\rm log}\left( {\Lambda}/{\m} \right)}} {\tau}_i =  2(Nn_i + 1) 
- \sum_j a_{ij} N n_j =  2
$$
Clearly:
$$
{d\over{d{\rm log}\left( {\Lambda}/{\m} \right)}} \sum_{k=0}^{r}  n_k \tau_k
=  \left( -\sum_j a_{ij} n_{j} + 2n_i \right) = 0
$$
On the other hand, from \cplngs\ we see that
$$
\tau = {1\over{\vert \Gamma \vert}} \sum_k n_k \tau_k
$$
and we come to the  complete agreement with the space-time picture
of the dilaton being constant.

{\sl Instanton corrections.} As the space-time dilaton
is constant for the solutions we discuss we may hope that the solutions we
find are valid, at least far away from the locations
of the fractional branes, but not too far to make
some of the gauge factors extremely strongly coupled. 
When we approach the branes,
the ${\rm log}(z)$ behaviour ${\g}$ makes the
fractional D-instantons favorable. The creation
amplitude  of the fractional 
D-instanton of the  ${\CR}_j$ type is clearly of the order of
\eqn\crea{\sim \exp 2\pi i {\g}_j \to \infty} near the
fractional D3-brane. The fractional $(p,q)$  D-instantons are
the 
euclidean $(p,q)$-string world sheets
wrapped around the collapsed two cycles $C_j$.
The field theory counterpart
of the picture which we are advocating is the possibility of 
having instanton corrections coming from the instantons
of each group factor. Also, the appearance of the $(p,q)$ types
of the instantons is the consequence of the S-duality
non-trivially realized in the field theories under consideration.

\bigskip
\bigskip
\noindent
{\bf Acknowledgements}
\bigskip
We are grateful to O.~Bergman, C.~Johnson,
J.~Minahan, A.~Polyakov and S.~Shatashvili
for useful discussions. 
The work of I.~R.~K. was supported in part by the NSF
grant PHY-9802484 and by the James S. McDonnell
Foundation Grant No. 91-48;
that of N.~A.~N 
by Robert H.~Dicke Fellowship from Princeton University,
partly by
RFFI under grant 98-01-00327,
partly  by the grant
96-15-96455 for scientific schools.


\vfill\eject
\listrefs
\end